\documentclass[letterpaper,preprintnumbers,prd,twocolumn,nofootinbib,nobibnotes,showpacs]{revtex4}
\usepackage{amsfonts}
\usepackage{mathrsfs}
\usepackage{epsfig}
\usepackage{graphicx}%
\usepackage{dcolumn}
\usepackage{amsmath}

\makeatletter
\def\btt#1{\texttt{\@backslashchar#1}}%
\DeclareRobustCommand\bblash{\btt{\@backslashchar}}%
\makeatother
\begin{document}

\title{Generalized modified gravity with the second order acceleration equation}
\author{Changjun Gao}\email{gaocj@bao.ac.cn}\affiliation{$^{}$The
National Astronomical Observatories, Chinese Academy of Sciences,
Beijing 100012, China}

\date{\today}

\begin{abstract}
In the theories of generalized modified gravity, the acceleration
equation is generally fourth order. So it is hard to analyze the
evolution of the Universe. In this paper, we present a class of
generalized modified gravity theories which have the acceleration
equation of second order derivative. Then both the cosmic
evolution and the weak-field limit of the theories are easily
investigated. We find that not only the Big-bang singularity
problem but also the current cosmic acceleration problem could be
easily dealt with.
\end{abstract}


\maketitle

\section{motivation}
In history, the motivation for modifying GR (General Relativity)
mainly comes from the fact that GR is not renormalizable. So it
can not be conventionally quantized. In the first place, Utiyama
and DeWitt showed that the renormalization at one-loop requires
the higher order curvature terms in the action of gravity theories
\cite{uti:62}. Secondly, Stelle showed the corresponding gravity
theories with these higher order terms are indeed renormalizable
\cite{ste:77}. Finally, when quantum effects or string theory are
taken into account, the effective low energy gravitational action
also requires higher order curvature invariants
\cite{bir:82,buc:92,vil:92}. So it was generally believed that the
modifications to GR would be important only at the scales of very
close to the Planck lengthy or Planck energy. Consequently, both
the Big-bang singularity and black hole singularity are expected
to be absent in the modified gravity theories
\cite{sta:80,bra:92,bra:93,muk:92,sha:90,tro:93}. This is the
belief before 1998.

However, with the discovery of cosmic acceleration in 1998
\cite{per:99,rie:98}, one realize that GR may also need to be
modified on very large scale or at very low energy (or very weak
gravitational field). These constitute the infrared modifications
to GR, for example, the GDP (Dvali-Gabadadze-Porrati) model
\cite{dva:00}, the $1/R$ modified gravity model \cite{car:04} and
so on. Here we shall not produce an exhaustive list of references
on modified gravity, but we prefer the readers to read the nice
review paper by Sotiriou and Faraoni \cite{faraoni:08} and the
references~therein. In general, the equations of motion for the
generalized modified gravity are of fourth order and one can
expect the particle content of the theory would have eight degrees
of freedom: two for the massless graviton, one in a scalar
excitation and five in a ghost-like massive spin two field
\cite{hin:95}. The presence of ghost leads one to accept
unphysical negative energy states in the theory and the property
of unitarity is lost \cite{haw:02}. This ghost problem is closely
related to the higher order property of the theories.

So the purpose of this paper is to seek for the \emph{second
order} theories of gravity, at least in the background of
spatially flat FRW (Friedmann-Robertson-Walker) Universe. Except
for satisfying the requirement of second-order, the theories also
meet ghost free conditions. Due to the property of second order of
acceleration equation, the resulting Friedmann equation remains
first order and the cosmic evolution of the universe is easily
deal with.

The paper is organized as follows. In section II, we briefly
review the generalized modified gravity theories. The equations of
motion are presented. In section III, we propose the Lagrangian
for the generalized modified gravity which are both ghost free and
second order (in the background of spatially flat FRW Universe).
In section IV, we investigate the cosmic evolution of some
specific models of Lagrangian. In section V, we investigate the
weak field limit of these models. Section IV gives the conclusion
and discussion.

We shall use the system of units in which $G=c=\hbar=1$ and the
metric signature $(-,\ +,\ +,\ +)$ throughout the paper.

\section{generalized modified gravity }
The generalized modified gravity theories have the action of the
form \cite{mad:89}
\begin{eqnarray}
\label{eq:mod} S&=&\int
d^{4}x\sqrt{-g}\left\{\frac{1}{16\pi}\left[{R}+f\left(R,\ P,\
Q\right)\right]+\mathscr{L}_{m}\right\}\;,
\end{eqnarray}
where $f$ is a general function of the Ricci scalar $R$ and two
curvature invariants,
\begin{eqnarray}
\label{eq:PQ}
 P\equiv R_{\mu\nu}R^{\mu\nu}\;,\ \ \
 Q\equiv R_{\mu\nu\sigma\lambda}R^{\mu\nu\sigma\lambda}\;,
\end{eqnarray}
which are of lowest mass dimension and parity-conserving.
$R_{\mu\nu}$ and $R_{\mu\nu\sigma\lambda}$ are the Ricci tensor
and the Riemann tensor, respectively. $\mathscr{L}_{m}$ is the
Lagrangian density for matters.

If we define

\begin{eqnarray}
 f_R\equiv \frac{\partial f}{\partial R}\;,\ \ \
 f_P\equiv \frac{\partial f}{\partial P}\;,\ \ \
 f_Q\equiv \frac{\partial f}{\partial Q}\;,
\end{eqnarray}
then we obtain the generalized Einstein equations \cite{mad:89}

\begin{eqnarray}
\label{Einstein}
 &&\mathscr{G}_{\mu\nu}\equiv G_{\mu\nu}-\frac{1}{2}g_{\mu\nu}f+f_RR_{\mu\nu}+2f_PR^{\lambda}_{\ \ \mu}R_{\lambda\nu}
 \nonumber\\&&+2f_QR_{\lambda\sigma\delta\mu}R^{\lambda\sigma\delta}_{\ \ \ \ \nu}
 +g_{\mu\nu}\nabla^2 f_{R}-\nabla_{\mu}\nabla_{\nu}f_{R}
 \nonumber\\&&-2\nabla_{\lambda}\nabla_{\sigma}\left[f_{P}R^{\lambda}_{\ \ (\mu}\delta^{\sigma}_{\ \ \nu)}\right]
 +\nabla^2\left(f_{P}R_{\mu\nu}\right)\nonumber\\&&+g_{\mu\nu}\nabla_{\lambda}\nabla_{\sigma}\left(f_PR^{\lambda\sigma}\right)
 -4\nabla_{\lambda}\nabla_{\sigma}\left[f_QR^{\lambda \ \ \ \ \ \sigma}_{\ \ (\mu\nu)}\right]\nonumber\\&&
 =8\pi T_{\mu\nu}\;.
\end{eqnarray}
$\mathscr{G}_{\mu\nu}$ is the generalized Einstein tensor,
$G_{\mu\nu}$. The same as the Einstein tensor, it satisfies the
Bianchi identity
\begin{eqnarray}
\mathscr{G}_{\mu\nu}^{\ \ ;\nu}=0\;.
\end{eqnarray}
But different from the Einstein tensor which is up to second order
derivative, $\mathscr{G}_{\mu\nu}$ is up to fourth-order. So
Eqs.~(\ref{Einstein}) are usually a set of fourth-order
differential equations except for the case of a cosmological
constant $f=2\Lambda$. The existence of higher derivatives in the
equation of motions suggests that one would always find ghosts in
a linearized analysis. Actually, it can be argued by considering
the Cauchy problem, the gauge symmetries and constraints on the
theory (see for instance \cite{hin:95}) that the generalized
modified gravity will contain at most eight degrees of freedom:
two for the usual massless graviton, one for a scalar field and
five for ghost-like massive spin two excitation. The ghost problem
leads one to accept negative energy states in the theory. So the
property of unitarity is lost \cite{haw:02}.

However, Comelli \cite{come:05}, Navarro and Acoleyen
\cite{nav:06} showed that with a suitable choice of parameters,
the theory would be ghost-free. Actually, they showed that the
general Lagrangian of the form $\mathscr{L}=\mathscr{L}(R,\ P-4Q)$
are ghost free. So in this case it is left with only an extra
scalar degree of freedom to the gravitational sector. In the next
section, we shall seek for the Lagrangian which gives the
acceleration equation of second order derivative and the
corresponding theories are ghost free.

\section{The Lagrangian}
The spatially flat FRW metric is given by
\begin{eqnarray}
\label{eq:frw}
ds^2=-dt^2+a\left(t\right)^2\left(dr^2+r^2d\Omega^2\right)\;,
\end{eqnarray}
where $a(t)$ is the scale factor. Given the metric, we could
calculate the Ricci scalar $R$ and the curvature invariants, $P,\
Q$,
\begin{eqnarray}
\label{eq:RPQ} R&=&6\left(\dot{H}+2H^2\right)\;,\nonumber\\
P&=&12\left[\left(\dot{H}+H^2\right)^2+H^2\left(H^2+\dot{H}\right)+H^4\right]\;,\nonumber\\
Q&=&12\left[\left(\dot{H}+H^2\right)^2+H^4\right]\;,
\end{eqnarray}
where $H\equiv{\dot{a}}/{a}$ is the Hubble parameter and dot
denotes the derivative with respect to the cosmic time $t$.

The second derivative of the scale factor, $\ddot{a}$ (in
$\dot{H}$), is present in $R,\ P,\ Q$. If $f\propto R$, the
corresponding equations of motion are the Einstein equations. They
are second-order differential equations. In this scenario,
$\dot{H}$ appears linearly in the Lagrangian. However, if
$\dot{H}$ appears nonlinearly in the Lagrangian just as
contributed by $R^2, \ P$ and $Q$, the corresponding equations of
motion would be fourth-order differential equations.

It is not hard to conjecture that, if we are able to make
$\dot{H}$ disappear in the Lagrangian $f$ such that it is uniquely
the function of Hubble parameter $H$, the resulting equation of
motion must be of second-order differential equation. Then how to
make $\dot{H}$ disappear? We can examine the proper combination of
$R^2,\ P,\ Q$.

To this end, let us calculate
\begin{eqnarray}
I&\equiv&\alpha R^2+\beta P+\gamma
Q=12\left(3\alpha+\beta+\gamma\right)\dot{H}^2\nonumber\\
&&+12\left(12\alpha+3\beta+2\gamma\right)\left(H^2\dot{H}+{H}^4\right)\;.
\end{eqnarray}
So if we let
\begin{eqnarray}
12\alpha+3\beta+2\gamma=0\;,
\end{eqnarray}
namely,
\begin{eqnarray}\label{eq:gamma}
\gamma=-6\alpha-\frac{3}{2}\beta\;,
\end{eqnarray}
we would obtain

\begin{eqnarray}
I=-\left(36\alpha+6\beta\right)\dot{H}^2\;.
\end{eqnarray}
Is $\dot{H}$ negative or positive? In order to answer this
question, let us resort to the Einstein equation in FRW Universe
\begin{eqnarray}
\dot{H}\propto-\left(\rho+p\right)\;,
\end{eqnarray}
where $\rho,\ p$ are the total cosmic energy density and pressure,
respectively. It is apparent $\dot{H}$ is nonpositive in the
history of the Universe which mainly covers three epochs dominated
by radiation, matter and cosmological constant, respectively.
Therefore, we assume
\begin{eqnarray}
\dot{H}\leq 0\;,
\end{eqnarray}
in the following. Thus
\begin{eqnarray}
\sqrt{I}=\sqrt{-36\alpha-6\beta}\left(-\dot{H}\right)\;.
\end{eqnarray}
It is apparent $\alpha$ and $\beta$ should obey
\begin{eqnarray}
-36\alpha-6\beta\geq 0\;.
\end{eqnarray}
If we define $J$ as follows
\begin{eqnarray}
J&\equiv&\frac{1}{\sqrt{12}}\sqrt{R+\frac{6}{\sqrt{-36\alpha-6\beta}}\sqrt{\alpha
R^2+\beta P+\gamma Q}}\;,\nonumber\\
\end{eqnarray}
then we have
\begin{equation}
J=H\;,\nonumber\\
\end{equation}
in the background of FRW Universe.

Now we could conclude that, for the general function of $f(J)$,
the Lagrangian density
\begin{eqnarray}
\label{eq:LLL}
&&\mathscr{L}=\frac{1}{16\pi}\left[R+f\left(J\right)\right]+\mathscr{L}_m\;,
\end{eqnarray}
always leads to the second order equation of motion in the
background of spatially flat FRW Universe. Of course, we have
assumed the energy-momentum tensor contributed by the matters is
up to second-order.

In general, the above theories of gravity would contain the
massive spin two ghost field in addition to the usual massless
graviton and the massive scalar \cite{faraoni:08}. But the $f(R)$
theories of gravity are found to be ghost free.
Ref.~\cite{buch:92} and Refs. \cite{come:05,nav:06} showed that
the models given by
\begin{eqnarray}\label{eq:LLLLL}
&&\mathscr{L}=\frac{1}{16\pi}\left[R+f\left(R,\
4P-Q\right)\right]\;,
\end{eqnarray}
are also ghost free. In view of this point, we should let
\begin{eqnarray}
\frac{\beta}{\gamma}=-4\;.
\end{eqnarray}
Taking account of Eq.~(\ref{eq:gamma}), we have

\begin{eqnarray}
\beta=-\frac{24}{5}\alpha\;,\ \ \ \gamma=\frac{6}{5}\alpha\;.
\end{eqnarray}

So $J$ is found to be

\begin{eqnarray}
&&J=\frac{1}{\sqrt{12}}\sqrt{{R+\sqrt{6\left(4P-Q\right)-5R^2}}}\;.
\end{eqnarray}
Using the Gauss-Bonnet invariant:

\begin{eqnarray}
&&G=R^2-4P+Q\;,
\end{eqnarray}
we have
\begin{eqnarray}\label{eq:JJ}
&&J=\frac{1}{\sqrt{12}}\sqrt{{R+\sqrt{R^2-6G}}}\;.
\end{eqnarray}
Then we recognize that the Lagrangian
\begin{eqnarray}\label{eq:LLLLLL}
&&\mathscr{L}=\frac{1}{16\pi}\left[R+f\left(J\right)\right]\;,
\end{eqnarray}
is actually the modified Gauss-Bonnet gravity which has been
studied in Ref.~\cite{cog:06}. In general, the acceleration
equation in the modified Gauss-Bonnet gravity is four order.
However, for our specific case (Eq.~(\ref{eq:JJ})), we would
obtain a second order acceleration equation from the Lagrangian
Eq.~(\ref{eq:LLLLLL}). We note that, in Eq.~(\ref{eq:LLLLLL}), $J$
is the function of $R, P, Q$ or $R, G$ according to
Eq.~(\ref{eq:JJ}). In the background of four dimensional FRW
Universe, we have $J=H$. But it is not the case for other
spacetimes.

Here we would like to point out that the above conclusion is valid
only for four dimensional FRW Universe. If the dimension of
spacetime is greater than four, the variation of
Eq.~(\ref{eq:LLLLLL}) would lead to a fourth order acceleration
equation. The reason could be understood as follows. In the
gauss-Bonnet gravity, if the dimension of spacetime is four, the
Gauss-Bonnet term turns out to be a topological invariant and so
it makes no contribution to the equation of motion. In higher
dimensions, the Gauss-Bonnet term is not a topologically invariant
and so it would contribute to the equation of motion. For the
modified Gauss-Bonnet gravity, the equation of motion is usually
fourth order even in the four dimensions.

Actually, the Gauss-Bonnet term is generalized in the Lovelock
gravity theory \cite{love:71}. It is found that the equations of
motion in Lovelock gravity are of second order in any spacetime
with any dimensions, needless to say in the background of four
dimensional FRW Universe. However, there is a notable difference
between the Lovelock gravity and the generalized modified gravity
in Eq.~(\ref{eq:LLLLLL}). The Lovelock gravity is an ultraviolet
modification to General relativity and unable to achieve the
infrared modification to gravity.

\section{second-order acceleration equation and the Friedmann equation}
In this section, we shall derive the acceleration equation and the
Friedmann equation from the Lagrangian density Eq.~(\ref{eq:LLL})
or Eq.~(\ref{eq:LLLLLL}). In the background of spatially flat FRW
Universe, the Lagrangian function is given by
\begin{eqnarray}
\label{eq:LLLL}
{L}=a^3\left\{\frac{1}{16\pi}\left[R+f\left(J\right)\right]+\mathscr{L}_m\right\}\;.
\end{eqnarray}
Substituting the Lagrangian function into the Euler-Lagrange
equation

\begin{eqnarray}
\frac{d^2}{dt^2}\left(\frac{\partial
L}{\partial\ddot{a}}\right)-\frac{d}{dt}\left(\frac{\partial
L}{\partial\dot{a}}\right)+\frac{\partial L}{\partial{a}}=0\;,
\end{eqnarray}
we obtain the acceleration equation

\begin{eqnarray}\label{eq:acc}
2\dot{H}+3H^2-\frac{1}{2}Hf^{'}+\frac{1}{2}f-\frac{1}{6}\dot{H}f^{''}=-8\pi
p\;,
\end{eqnarray}
where the prime denotes the derivative with respect to $H$. It is
obvious the acceleration equation belongs to the second-order
differential equations.

Using the acceleration equation, we could obtain the Friedmann
equation from the energy-conservation equation

\begin{eqnarray}\label{eq:ece}
\frac{d\rho}{dt}+3H\left(\rho+p\right)=0\;,
\end{eqnarray}
as follows
\begin{eqnarray}\label{eq:Fried}
3H^2+\frac{1}{2}f-\frac{1}{2}Hf^{'}=8\pi\rho\;.
\end{eqnarray}
The left hand side of the Friedmann equation is only the function
of Hubble parameter. It is remarkably simple in the investigation
of the evolution of the Universe.

Eqs.~(\ref{eq:acc},\ \ref{eq:ece},\ \ref{eq:Fried}) and the
equation of state for matters
\begin{eqnarray}\label{eq:eos}
p=p\left(\rho\right)\;.
\end{eqnarray}
constitute the main equations which govern the evolution of the
universe. Among the four equations, only three of them are
independent. For convenience, we always focus on
Eqs.~(\ref{eq:ece},\ \ref{eq:Fried},\ \ref{eq:eos}).

\section{some examples}
In this section, we shall study some specific and interesting
forms of $f(H)$.
\subsection{$\Lambda \textrm{CDM}$ Model}
If $f=-16\pi\Lambda $ with $\Lambda$ a constant, we obtain from
Eq.~(\ref{eq:Fried})
\begin{eqnarray}
3H^2=8\pi\left(\rho+\Lambda\right)\;.
\end{eqnarray}
This is the Friedmann equation for $\Lambda \textrm{CDM}$
($\Lambda$-Cold Dark Matter) universe. Although the $\Lambda
\textrm{CDM}$ model provides an excellent fit to the wealth of
high-precision observational data, on the basis of a remarkably
small number of cosmological parameters \cite{dun:08}, it is
plagued with the well-known cosmological constant problem and the
cosmic coincidence problem which prompted cosmologists to look for
other explanation for the observed accelerated expansion.
\subsection{Power law for $f$}

Assume the energy density contributed by $f$ is the form of a
power law $\eta H^{n}$ with $\eta$ a positive constant and $n$ an
integer. Then from the Friedmann equation Eq.~(\ref{eq:Fried}) we
have
\begin{eqnarray}
-\frac{1}{2}f+\frac{1}{2}Hf^{'}=8\pi\eta H^{n}\;.
\end{eqnarray}
Thus $f$ is derived as
\begin{eqnarray}
f=16\pi\eta \frac{H^{n}}{n-1}\;.
\end{eqnarray}
When $n=1$, we have
\begin{eqnarray}
f=16\pi\eta H\ln{H}\;.
\end{eqnarray}
Now we have the conclusions as follows.

$\textbf{(1)}.$ When $f=16\pi\eta J\ln{J}$ (for $n=1$), we have
the following Friedmann equation

\begin{eqnarray}
3H^2=8\pi\left(\rho+\eta H\right)\;.
\end{eqnarray}

It is the same as the Friedmann equation given by the DGP modified
gravity \cite{dgp:00}. The equation can be rewritten as
\begin{eqnarray}
3H^2=8\pi\rho+\frac{1}{6}\eta^2+\frac{1}{6}\eta\sqrt{\eta^2+96\pi\rho}\;.
\end{eqnarray}
 If we define

\begin{eqnarray}
\eta\equiv \sqrt{6\rho_I/\pi}/4\;,
\end{eqnarray}
we have

\begin{eqnarray}
3H^2=8\pi\left[\rho+\frac{\rho_I}{2}+\sqrt{\frac{\rho_I}{2}\left(2\rho+\frac{\rho_I}{2}\right)}\right]\;.
\end{eqnarray}
Here $\rho_I$ is a constant energy density. The $\rho_I$ terms are
investigated as the candidate of dark energy in many literatures,
for example, \cite{dgpref} and references therein. Different from
the cosmological constant, this dark energy density increases with
the increasing of background energy density $\rho$. So the cosmic
coincidence problem is greatly relaxed. But it is argued that the
DGP model is disfavored by the history of cosmic structure
formation \cite{fang:08} because of the fast increasing of dark
energy density with redshifts.

$\textbf{(2)}.$ When $f=-\eta J^{-2}$ (for $n=-2$) with $\eta$ a
constant, we obtain from Eq.~(\ref{eq:Fried})
\begin{eqnarray}
3H^2-\frac{3}{2}\eta H^{-2}=8\pi\rho\;.
\end{eqnarray}
The above equation can be rewritten as
\begin{eqnarray}\label{eq:s}
3H^2=4\pi\rho+\sqrt{16\pi^2\rho^2+\frac{9}{2}\eta}\;.
\end{eqnarray}
If we define
\begin{eqnarray}
\eta\equiv\frac{128}{9}\pi^2\rho_I^2\;,
\end{eqnarray}
then Eq.~(\ref{eq:s}) can be rewritten as

\begin{eqnarray}
3H^2=4\pi\left(\rho+\sqrt{\rho^2+4\rho_I^2}\right)\;.
\end{eqnarray}
It is apparent $\rho_I$ plays the role of a constant energy
density. When $\rho\gg\rho_I$, it restores to the standard
Friedmann equation. When $\rho\ll\rho_I$, the Universe evolves
into a de Sitter phase. We have shown that this model could
interpret the current acceleration of the Universe \cite{gao:09}.
Different from the DGP model, the dark energy density in this
scenario decreases with the increasing of redshifts. So the cosmic
coincidence problem is also relaxed.

 $\textbf{(3)}.$ When $f=16\pi\eta {J^{4}}/{3}$ (for $n=4$), we have the Friedmann equation as
follows

\begin{eqnarray}
3H^2=8\pi\left(\rho+\eta H^4\right)\;.
\end{eqnarray}
It is a quadratic equation of $H^2$. Mathematically, we would have
two roots for $H^2$. But physically, only one of which could
reduce to the standard Friedmann equation in the limit of small
$\rho$. So the physical root takes the form of

\begin{eqnarray}\label{eq:RS}
3H^2=\frac{3}{16\pi\eta}\left(3-\sqrt{9-256\pi^2\rho\eta}\right)\;.
\end{eqnarray}

\begin{figure}
\includegraphics[width=8.5cm]{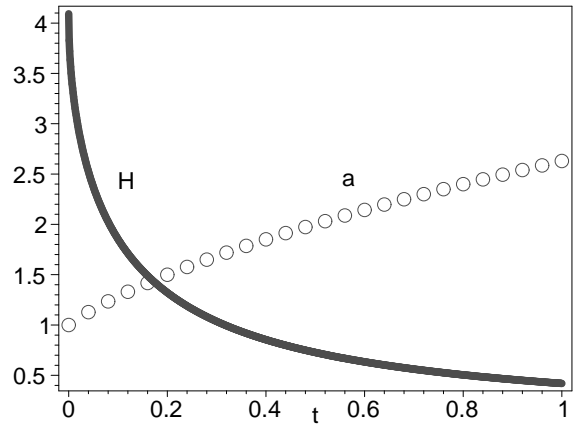}
\\
\caption{The evolution of the scale factor $a$ and the Hubble
parameter $H$ with respect to the cosmic time $t$. It shows that
the universe is created in finite time with finite scale factor
and finite Hubble parameter. So the Big-Bang singularity is
avoided.} \label{fig:ah}
\end{figure}

Define

\begin{eqnarray}
\eta\equiv \frac{9}{128\pi^2\rho_U}\;,
\end{eqnarray}
with $\rho_U$ some positive constant. The Friedmann equation
Eq.~(\ref{eq:RS}) turns out to be

\begin{eqnarray}\label{eq:RSRS}
3H^2={8\pi\rho_U}\left(1-\sqrt{1-\frac{2\rho}{\rho_U}}\right)\;.
\end{eqnarray}
Here $\rho_U$ plays the role of a constant energy density. It is
apparent $\rho$ should obey $\rho\leq\rho_U$. So, to the zeroth
order of $\rho/\rho_U$, we obtain the standard Friedmann equation.
To the first order of $\rho/\rho_U$, we obtain the Friedmann
equation in the Randall-Sundrum brane world model \cite{RS:99}

\begin{eqnarray}
3H^2=8\pi\left(\rho+\frac{\rho^2}{2\rho_U}\right)\;.
\end{eqnarray}
Putting
\begin{eqnarray}\label{eq:rrrrrr}
\rho_U=1\;, \ \ \rho=\frac{1}{a^4}\;,
\end{eqnarray}
and using Eq.~(\ref{eq:RSRS}), we plot the evolution of the scale
factor $a$ and the Hubble parameter $H$ in Fig.~\ref{fig:ah},
respectively. It shows that the universe is created in finite time
with finite scale factor and finite Hubble parameter. So the
Big-Bang singularity is avoided.

It is apparent the energy density is also finite from
Eq.~(\ref{eq:RSRS}). Then how about the pressure and its
higher-derivatives? If they are irregular, some weak singularities
would appear. We find that there are regular and there are no weak
singularities. The proof are as follows. Taking account of
Eq.~(\ref{eq:rrrrrr}) and the energy conservation equation, we
obtain the pressure $p=\rho/3$. Since $\rho$ is regular, $p$ is
also regular. Then with the help of energy conservation equation,
we find the higher derivatives of pressure are regular.

\subsection{Past de Sitter universe }

In the last part of subsection $\textrm{B}$, we find the universe
could be created in finite time with finite scale factor and
finite Hubble parameter. So it seems there exists a starting point
of time. Different from this case, in this section, we present a
past de Sitter universe. In this scenario, the universe starts
from a de Sitter phase. So the history of the universe is
infinite. In other words, there does not exist a starting point of
cosmic time.

In order to construct a past de Sitter universe, we could explore
\begin{eqnarray}
f=6J^2-4J\sqrt{{6\pi}{\rho_U}}\tanh^{-1}\left({\sqrt{\frac{3J^2}{8\pi\rho_U}}}\right)\;,
\end{eqnarray}
where $\rho_U$ is a constant energy density. Substituting it into
Eq.~(\ref{eq:Fried}), we obtain
\begin{eqnarray}\label{eq:UV0}
3H^2=\left(1-\frac{3H^2}{8\pi\rho_U}\right)8\pi{\rho}\;,
\end{eqnarray}
or
\begin{eqnarray}\label{eq:UV}
3H^2=8\pi\frac{\rho}{1+{\rho}/{\rho_U}}\;.
\end{eqnarray}
To the zero order of ${\rho}/{\rho_U}\ll 1$, it restores to the
Friedmann equation in General Relativity. To the first order of
${\rho}/{\rho_U}$, we have

\begin{eqnarray}
3H^2=8\pi\left(\rho-\frac{\rho^2}{\rho_U}\right)\;.
\end{eqnarray}
It is the Friedmann equation in the theory of loop quantum gravity
\cite{loop:11,loop:22}

On the other hand, when ${\rho}/{\rho_U}\gg 1$, we obtain
\begin{eqnarray}
3H^2=8\pi{\rho_U}\;.
\end{eqnarray}
This is for de Sitter universe. Taking into account all matter
sources which include relativistic matter (radiation), baryon
matter, dark matter and the cosmological constant, we can rewrite
the Friedmann equation Eq.~(\ref{eq:UV}) as follows
\begin{eqnarray}
h^2=\frac{\frac{\Omega_{r0}}{a^4}+\frac{\Omega_{m0}}{a^3}+\Omega_{\lambda}}
{1+\frac{\frac{\Omega_{r0}}{a^4}+\frac{\Omega_{m0}}{a^3}+\Omega_{\lambda}}{\Omega_U}}\;.
\end{eqnarray}
Observations show that $\Omega_{r0}=8.1\cdot 10^{-5}$,
$\Omega_{m0}=0.27$, $\Omega_{\lambda}=0.73$, which are the ratio
of energy density for radiation, matter (including baryon matter
and dark matter) and cosmological constant in the present-day
universe. The dimensionless Hubble parameter $h$ is defined by
\begin{eqnarray}
h=\frac{H}{H_0}\;,
\end{eqnarray}
with $H_0$ the present-day Hubble parameter. $\Omega_{U}$ is the
dimensionless energy density contributed by $\rho_U$.

\begin{figure}
\includegraphics[width=8.5cm]{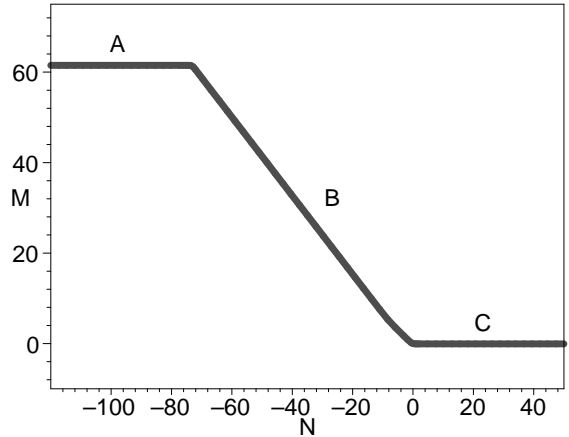}
\\
\caption{There are three epochs $A, B, C$ in the total life of the
Universe. The epoch of $A$ corresponds to the de Sitter phase. The
Universe exponentially expands (inflating) in this period. Then
the inflation stops around the redshift of $z\sim 10^{30}$. The
epoch of $B$ is dominated firstly by the radiation and then by the
matter. It stops around the redshift of $z\sim 0$. The epoch of
$C$ is dominated by a small cosmological constant.  ${\textrm{C}}$
is the future de Sitter phase.} \label{fig:evolution}
\end{figure}

Taking $\Omega_U=10^{123}$ (which represents $\rho_U$ is Planck
energy density), we plot the evolution of the re-scaled
dimensionless Hubble parameter $M\equiv\log_{10} h$ with respect
to the re-scaled scale factor $N\equiv\ln a$ in
Fig.~(\ref{fig:evolution}). There are three epochs $A, B, C$ in
the total life of the Universe. The epoch of $A$ corresponds to
the de Sitter phase. The Universe exponentially expands
(inflating) in this period. Then the inflation stops around the
redshift of $z\sim 10^{30}$. The epoch of $B$ is dominated firstly
by the radiation and then by the matter. It stops around the
redshift of $z\sim 0$. The epoch of $C$ is dominated by a small
cosmological constant. It is the future de Sitter phase.

\section{weak field limit}
In this section, we shall study the weak field limit of the above
modified gravity theories. Following Ref. \cite{come:05,nav:06},
we expand the action in powers of the curvature perturbations.
Then it can be shown that at the bilinear level the linearisation
of the theory over a maximally symmetric spacetime will be the
same as the theory obtained in \cite{hin:95,chi:05},

\begin{eqnarray}
S&=&\int d^4x\sqrt{-g}\frac{1}{{16\pi}}\left(R-\Lambda+\delta
R+\frac{1}{6m_0^2}R^2\right.\nonumber\\&&\left.-\frac{1}{2m_2^2}C_{\mu\nu\sigma\lambda}C^{\mu\nu\sigma\lambda}\right)\;,
\end{eqnarray}
where $C_{\mu\nu\sigma\lambda}$ is the Weyl tensor and we have
defined

\begin{eqnarray}
\Lambda &\equiv& \textsc{<}f-Rf_R +
R^2\left(f_{RR}/2-f_P/4-f_Q/6\right)\nonumber\\&&+R^3
\left(f_{RP}/2 +
f_{RQ}/3\right)\nonumber\\&&+R^4\left(f_{PP}/8+f_{QQ}/18 +
f_{PQ}/6\right)>_0\;, \\
\delta &\equiv& \textsc{<}f_R-Rf_{RR}-
R^2\left(f_{RP}+2f_{RQ}/3\right)\nonumber\\&&-R^3 \left(f_{PP}/4
+ f_{QQ}/9+f_{PQ}/3\right)>_0\;, \\
m_0^{-2} &\equiv& \textsc{<}3f_{RR}+2f_{P}+2f_Q+
R\left(3f_{RP}+2f_{RQ}\right)\nonumber\\&&+R^2 \left(3f_{PP}/4 +
f_{QQ}/3+f_{PQ}\right)>_0\;, \\
m_2^{-2} &\equiv& -\textsc{<}f_P+4f_Q>_0\;.
\end{eqnarray}
Here $<\cdot\cdot\cdot>_0$ represent the values of the
corresponding quantities in the background of some spacetime and
$F_{RR}\equiv\frac{\partial^2 f}{\partial R^2}$, etc. It is
apparent for the action, Eq.~(\ref{eq:LLLLLL}), the inverse of
mass squared of the ghost is $m_2^{-2}=0$. Thus there is no ghost
in the spectrum. But there is still an extra scalar with the mass
$m_0$. The $\Lambda$ term behaves as the vacuum energy and the
$\delta$ term contributes the variation of the gravitational
constant,
\begin{eqnarray}
\frac{\delta G}{G}=\frac{-\delta}{1+\delta}\simeq-\delta\;.
\end{eqnarray}
\begin{table*}[t]
\begin{center}
\begin{tabular}{|c|c|c|c|c|}
 \hline
 Models &  $\Lambda$ & $\delta$ & $m_0^{-2}$ & $m_2^{-2}$  \\
\hline \hline (a) & $\Lambda$ & 0 & $0$ & 0 \\
\hline (b) & $-\sqrt{\Lambda\rho_I}(9\ln 2+3\ln\pi+3\ln\Lambda-8-3\ln 3)/16$ & $-\sqrt{\rho_I/\Lambda}(-3\ln 3+4+9\ln 2+3\ln \pi+3\ln\Lambda)/32$ & $m_0=0$  & $0$ \\
\hline (c) & $\rho_I^2/\Lambda$ & $-{\rho_I^2}/{(4\Lambda^2)}$ & $m_0=0$ & $0$ \\
\hline (d) & $0$ & $0$ & $m_0=0$ & $0$
 \\
\hline (e) & $0$ & $0$ & $m_0=0$ & $0$
 \\\hline
\end{tabular}
\end{center}
\caption[crit]{The parameters of five models in the background of
de Sitter spacetime.} \label{par1}
\end{table*}
\begin{table*}[t]
\begin{center}
\begin{tabular}{|c|c|c|c|c|}
 \hline
 Models &  $\Lambda$ & $\delta$ & $m_0^{-2}$ & $m_2^{-2}$  \\
\hline \hline (a) & $0$ & 0 & $0$ & 0 \\
\hline (b) & $0$ & $0$ & $m_0=0$  & $0$ \\
\hline (c) & $+\infty$ & $-\infty$ & $m_0=0$ & $0$ \\
\hline (d) & $0$ & $0$ & $m_0=0$ & $0$
 \\
\hline (e) & $0$ & $0$ & $m_0=0$ & $0$
 \\\hline
\end{tabular}
\end{center}
\caption[crit]{The parameters of five models in the background of
Minkowski spacetime.} \label{par2}
\end{table*}
In TABLE.~$\textrm{I}$ and TABLE.~$\textrm{II}$, we calculate the parameters for the five models:\\
(a): $f=-16\pi\Lambda$\ ;\\
(b): $f=16\pi\eta J\ln{J}$ \ ;\ \\
(c): $f=-\eta J^{-2}$ \ ;\ \\
(d): $f=16\pi\eta {J^{4}}/{3}$\ ;\ \\
(e):
$f=6J^2-4J\sqrt{{6\pi}{\rho_U}}\tanh^{-1}\left({\sqrt{\frac{3J^2}{8\pi\rho_U}}}\right)$
\ ,\\
in the background of de Sitter spacetime and Minkowski spacetime
(by taking the limit of $\Lambda\longrightarrow 0$). From
TABLE~$\textrm{I}$, we see the models $a,b,c$ contribute the
non-vanishing cosmological constant terms. For models $d$ and $e$,
the $\Lambda$ terms are zero because they are the ultraviolet
modification to GR. In the column of $\delta$, we see the models
$b,c$ contribute the non-vanishing gravitational constant terms.
Except for the model of $a$, there exists a scalar degree of
freedom with vanishing mass in $b,c,d,e$.

In the background of Minkowski spacetime, we see from the
TABLE.~$\textrm{II}$ that, except for the model of $c$, the other
models make no contribution to vacuum energy and gravitational
constant. The reason for this is that the  model $c$ is
essentially a modification with inverse curvature invariants. So
the Minkowski spacetime does not solve the corresponding equations
of motion.

\section{Conclusion and discussion}
In the theories of generalized modified gravity, the acceleration
equation is generally fourth order. So it is hard to analyze the
evolution of the Universe \cite{car:04}. On the other hand, these
theories are also plagued with the ghost problem. So the property
of unitary of the theory is lost \cite{haw:02}. In view of this
point, we present a class of generalized modified gravity theories
which have the acceleration equation of second order derivative
and ghost free. Then we explore some specific examples for the
Lagrangian function. We find both the cosmic evolution and the
weak-field limit of the theories are easily investigated.
Furthermore, not only the Big-bang singularity problem but also
the current cosmic acceleration problem could be easily dealt
with.

\acknowledgments

We thank the referee for the expert and insightful comments, which
have certainly improved the paper significantly. This work is
supported by the National Science Foundation of China under the
Key Project Grant No. 10533010, Grant No. 10575004, Grant No.
10973014, and the 973 Project (No. 2010CB833004).

\newcommand\ARNPS[3]{~Ann. Rev. Nucl. Part. Sci.{\bf ~#1}, #2~ (#3)}
\newcommand\AL[3]{~Astron. Lett.{\bf ~#1}, #2~ (#3)}
\newcommand\AP[3]{~Astropart. Phys.{\bf ~#1}, #2~ (#3)}
\newcommand\AJ[3]{~Astron. J.{\bf ~#1}, #2~(#3)}
\newcommand\APJ[3]{~Astrophys. J.{\bf ~#1}, #2~ (#3)}
\newcommand\APJL[3]{~Astrophys. J. Lett. {\bf ~#1}, L#2~(#3)}
\newcommand\APJS[3]{~Astrophys. J. Suppl. Ser.{\bf ~#1}, #2~(#3)}
\newcommand\JHEP[3]{~JHEP.{\bf ~#1}, #2~(#3)}
\newcommand\JCAP[3]{~JCAP. {\bf ~#1}, #2~ (#3)}
\newcommand\LRR[3]{~Living Rev. Relativity. {\bf ~#1}, #2~ (#3)}
\newcommand\MNRAS[3]{~Mon. Not. R. Astron. Soc.{\bf ~#1}, #2~(#3)}
\newcommand\MNRASL[3]{~Mon. Not. R. Astron. Soc.{\bf ~#1}, L#2~(#3)}
\newcommand\NPB[3]{~Nucl. Phys. B{\bf ~#1}, #2~(#3)}
\newcommand\CQG[3]{~Class. Quant. Grav.{\bf ~#1}, #2~(#3)}
\newcommand\PLB[3]{~Phys. Lett. B{\bf ~#1}, #2~(#3)}
\newcommand\PRL[3]{~Phys. Rev. Lett.{\bf ~#1}, #2~(#3)}
\newcommand\PR[3]{~Phys. Rep.{\bf ~#1}, #2~(#3)}
\newcommand\PRD[3]{~Phys. Rev. D{\bf ~#1}, #2~(#3)}
\newcommand\RMP[3]{~Rev. Mod. Phys.{\bf ~#1}, #2~(#3)}
\newcommand\SJNP[3]{~Sov. J. Nucl. Phys.{\bf ~#1}, #2~(#3)}
\newcommand\ZPC[3]{~Z. Phys. C{\bf ~#1}, #2~(#3)}
 \newcommand\IJGMP[3]{~Int. J. Geom. Meth. Mod. Phys.{\bf ~#1}, #2~(#3)}
  \newcommand\GRG[3]{~Gen. Rel. Grav.{\bf ~#1}, #2~(#3)}

\end{document}